%% file: SolO_DFR.tex
\documentclass[preprint]{aastex701}

\begin{document}

\title{Probing Solar Wind Structures with Solar Energetic Particle Observations from Solar Orbiter}

\author[orcid=0000-0003-2865-1772,sname='Chen']{Xiaohang Chen}
\affiliation{State Key Laboratory of Lunar and Planetary Sciences, Macau University of Science and Technology, Macau 999078, China}
\email{xhchen@must.edu.mo}

\author[orcid=0000-0003-4695-8866,sname='Li']{Gang Li}
\affiliation{State Key Laboratory of Lunar and Planetary Sciences, Macau University of Science and Technology, Macau 999078, China}
\email[show]{gli@must.edu.mo}

\author[0000-0002-0850-4233,sname='Giacalone']{Joe Giacalone}
\affiliation{Lunar and Planetary Laboratory, University of Arizona, Tucson, AZ 85721, USA}
\email{giacalon@arizona.edu}

\author[orcid=0000-0003-1034-5857,sname='Kong']{Xiangliang Kong}
\affiliation{ Institute of Space Sciences, Shandong University, Shandong 264209, China}
\email{kongx@sdu.edu.cn}

\author[orcid=0000-0003-4245-3107,sname='Fu']{Shuai Fu}
\affiliation{State Key Laboratory of Lunar and Planetary Sciences, Macau University of Science and Technology, Macau 999078, China}
\email{sfu@must.edu.mo}  

\author[orcid=0009-0007-8896-4400,sname='Zhang']{Rumeng Zhang}
\affiliation{ Institute of Space Sciences, Shandong University, Shandong 264209, China}
\email{2961689783@qq.com}

\author[orcid=0009-0006-7646-8023,sname='Zhao']{Changyue Zhao}
\affiliation{ Institute of Space Sciences, Shandong University, Shandong 264209, China}
\email{15254803322@163.com}

\author[0000-0003-1093-2066,sname='Ho']{George C. Ho} 
\affiliation{Southwest Research Institute, San Antonio, TX 78238, USA}
\email{george.ho@swri.org}

\author[0000-0002-3176-8704,sname='Lario']{David Lario}
\affiliation{NASA Goddard Space Flight Center, Greenbelt, MD 20771, USA}
\email{david.larioloyo@nasa.gov}

\author[0000-0002-9829-3811,sname='Ding']{Zheyi Ding}
\affiliation{Institute of Experimental and Applied Physics, Kiel University, Leibnizstrasse 11, D-24118 Kiel, Germany}
\email{ding@physik.uni-kiel.de}

\author[0000-0002-7388-173X,sname='Wimmer-Schweingruber']{Robert F. Wimmer-Schweingruber}
\affiliation{Institute of Experimental and Applied Physics, Kiel University, Leibnizstrasse 11, D-24118 Kiel, Germany}
\email{wimmer@physik.uni-kiel.de}

\author[0000-0003-2169-9618,sname='Mason']{Glenn M. Mason}
\affiliation{Applied Physics Laboratory, Johns Hopkins University, Laurel, MD 20723, USA}
\email{Glenn.Mason@jhuapl.edu}

\author[0000-0002-2152-0115,sname='Bruno']{Roberto Bruno}
\affiliation{INAF—Istituto di Astrofisica e Planetologia Spaziali, Via Fosso del Cavaliere 100, 00133 Rome, Italy}
\email{roberto.bruno@inaf.it}

\author[0000-0001-6344-6956,sname='Wijsen']{Nicolas Wijsen}
\affiliation{Centre for Mathematical Plasma Astrophysics, KU Leuven Campus Kulak, 8500 Kortrijk, Belgium}
\email{nicolas.wijsen@kuleuven.be}

\author[0000-0002-4240-1115,sname='Rodríguez-Pacheco']{Javier Rodríguez-Pacheco}
\affiliation{Universidad de Alcalá, Space Research Group, Alcalá de Henares, 28801 Madrid, Spain}
\email{fsrodriguez@uah.es}

\input{./Tex1_Abstract}

\input{./Tex2_Introduction}

\input{./Tex3_Instrumentation}

\input{./Tex4_Observations}

\input{./Tex5_Conclusions}

\input{./Tex6_Acknowledgments.tex}

\clearpage
\bibliography{SolO_DFR}{}
\bibliographystyle{aasjournalv7}
\end{document}

%% file: Tex1_Abstract.tex
\begin{abstract}

The propagation of solar energetic particles (SEPs) through the heliosphere is primarily guided by the interplanetary magnetic field (IMF) which is embedded in the solar wind plasma. Large-scale IMF structures can drive transient variations in SEP intensities. Using Solar Orbiter observations, we identify a distinct class of SEP variations: SEP flux deflections (SFDs), which are commonly detected in SEP events and frequently recur multiple times within a single event. SFDs are characterized by a sudden change in SEP flux directions where the intensities drop in one direction and increase in another direction, without a significant net change in total flux magnitude. These deflections occur dispersionlessly across a broad energy range—from tens of keV to over 100 MeV—and exhibit steep intensity gradients. SFDs are typically associated with magnetic flux tubes with boundary features consistent with tangential discontinuities. We further show that the solar wind inside these structures exhibits distinct plasma properties, and that the SEP streaming direction within SFDs aligns closely to the flux-tube axis. These observations suggest that magnetic flux tubes are a prevalent structural element of the solar wind, and demonstrate that SEPs can serve as an effective diagnostic tool for probing the topology and dynamics of solar wind structures.

\end{abstract}

%% file: Tex2_Introduction.tex
\section{Introduction}
\label{sec:Introduction}

The propagation of solar energetic particles (SEPs) through the heliosphere is usually described as a stochastic process modulated by magnetic turbulence in the solar wind. Charged particles undergo parallel and perpendicular diffusion relative to the background magnetic field, which allows them to stream both along and across the interplanetary magnetic field (IMF) \citep[][]{Jokipii1966,Roelof1968,Giacalone1999,Matthaeus2003,Qin2004,Shalchi2010,Ruffolo2012}. SEP events usually exhibit spatial and temporal variations in the in-situ observations near the Earth which are believed to be related to a number of factors regarding the processes in SEP acceleration, transport and magnetic connection to the observers \citep{Guo2010,Giacalone2012,Giacalone2017,Lario2017,Kong2017,Chen2022,Liu2023,Wijsen2023}.

One of the most intriguing SEP phenomena is the transient variation in flux intensity known as dropouts \citep{Mazur2000,Gosling2004}. These events appear as sudden decreases in SEP intensities followed by rapid recoveries, with sharp and dispersionless boundaries across all energies. It has been proposed that dropouts arise from the intermittent magnetic disconnection between observers and SEP source regions caused by large-scale magnetic fluctuations \citep{Giacalone2000}. Dropouts are more often observed in impulsive SEP events than in large gradual events \citep{Chollet2008}. In this context, SEPs accelerated near the Sun propagate along magnetic field lines originating from a compact source region. Dropouts occur when the magnetic connection between the observer and the source region is temporarily disrupted. The boundaries of dropouts are so sharp that the characteristic length of the SEP gradient is observed to be on the order of the Larmor radius, indicating very weak cross-field diffusion \citep{Chollet2011}. An alternative mechanism has been proposed, in which these variations are produced by the trapping of energetic particles within topological structures in solar wind turbulence \citep{Ruffolo2003,Chuychai2007,Chuychai2005}. In this alternative picture, dropouts should typically be accompanied by structural changes in the solar wind and the magnetic field.

The recently launched Parker Solar Probe \citep[PSP;][]{Fox2016} and Solar Orbiter \citep[SolO;][]{Muller2020} missions provide unique opportunities to study the origin of these variations close to the Sun. Unlike classical dropout events, we observed a distinct class of SEP variations — referred to as SEP flux deflections (SFDs) — using the energetic particle observations from SolO. In SFDs, SEP fluxes suddenly decrease in one direction while simultaneously increasing in others, indicating a deflection rather than a dropout. These events also exhibit sharp, dispersionless boundaries, though with a smaller net change in the average intensities. Magnetic rotation is often observed during SEP deflections, indicating that the observer might be crossing a local magnetic structure in the solar wind. 

Coherent structures, such as magnetic flux tubes and current sheets, are considered a fundamental component of solar wind turbulence. These intermittent structures, which convect with the solar wind, are characterized by pressure-balanced and tangential discontinuities based on in-situ observations, with typical scales ranging from minutes to hours \citep{Tu1991,Tu1993,Bruno2001,Hu2004,Li2008,Miao.etal10}. Such structures could play a role in modulating the transport of SEPs \citep{Gosling2005,Qin2008,Li2011,Arnold2013} and have been found to be associated with dropout events \citep{Trenchi2013a,Trenchi2013b}. However, their origin  remains debated. On one hand, they may arise from the local generation within the solar wind via nonlinear turbulence cascades \citep{Farge1992,Chang2004,Zhou2004}. On the other hand, they could be entangled flux tubes originating at the solar surface, weaving together into spaghetti-like structures \citep{McCracken1966,Mariani1973,Neugebauer1981,Bruno2001,Borovsky2008}. Distinguishing between these two scenarios using single-point plasma measurements is challenging. In this context, SEPs could serve as a powerful probe to diagnose the origin of these structures.

In this study, we present observations of SFDs associated with crossings of magnetic flux tubes whose orientation differs from that of the background magnetic field. These observations suggest that magnetic flux tubes may be static and prevalent structural elements of the solar wind, offering new insights into its nature.

%% file: Tex3_Instrumentation.tex
\section{Instrumentation} \label{sec:Instrumentation}

SolO was launched in February 2020 into an elliptical orbit around the Sun with a perihelion distance decreasing to 0.28 au, offering a unique opportunity for in-situ observations of solar wind and energetic particles in the inner heliosphere.  For this study, we analyze energetic particle data from the Energetic Particle Detector suite \citep[EPD;][]{Rodriguez2020,Wimmer-Schweingruber2021}. EPD comprises four individual sensors: the Supra-Thermal Electron and Proton sensor (STEP), the Electron Proton Telescope (EPT), the High Energy Telescope (HET), and the Suprathermal Ion Spectrograph (SIS).  Specifically, we use Level 3 data from EPT at 1-minute cadence and 1-minute averages of Level 2 HET data. Together, EPT and HET measurements cover the energy range for protons and heavier ions from 25 keV nucleon$^{-1}$ to over 100 MeV nucleon$^{-1}$. The two telescopes are integrated into one unit comprising two double-ended detecters, each providing directional information in 4 different fields of view of incoming particles. EPT-HET 1 is oriented sunward and anti-sunward along the nominal Parker spiral, detecting particles coming from the Sun and those traveling toward the Sun, respectively. EPT-HET 2 points northward and southward, detecting particles arriving from the north and from the south, respectively. Magnetic field measurements are provided by the fluxgate vector magnetometer \citep[MAG;][]{Horbury2020}, which provides the magnetic field vector in the RTN coordinate system at a cadence of 8 vectors per second. The solar wind plasma data for the second event in 2023 (discussed in the following sections)—including proton density, temperature, and bulk speed—are obtained from the Proton and Alpha particle Sensor (PAS), part of the Solar Wind Analyser (SWA), at a 1-second cadence \citep{Owen2020}. 

%% file: Tex4_Observations.tex
\section{Observations} \label{sec:Observations}

\begin{figure}[ht]
\centering
\includegraphics[width=0.8\textwidth]{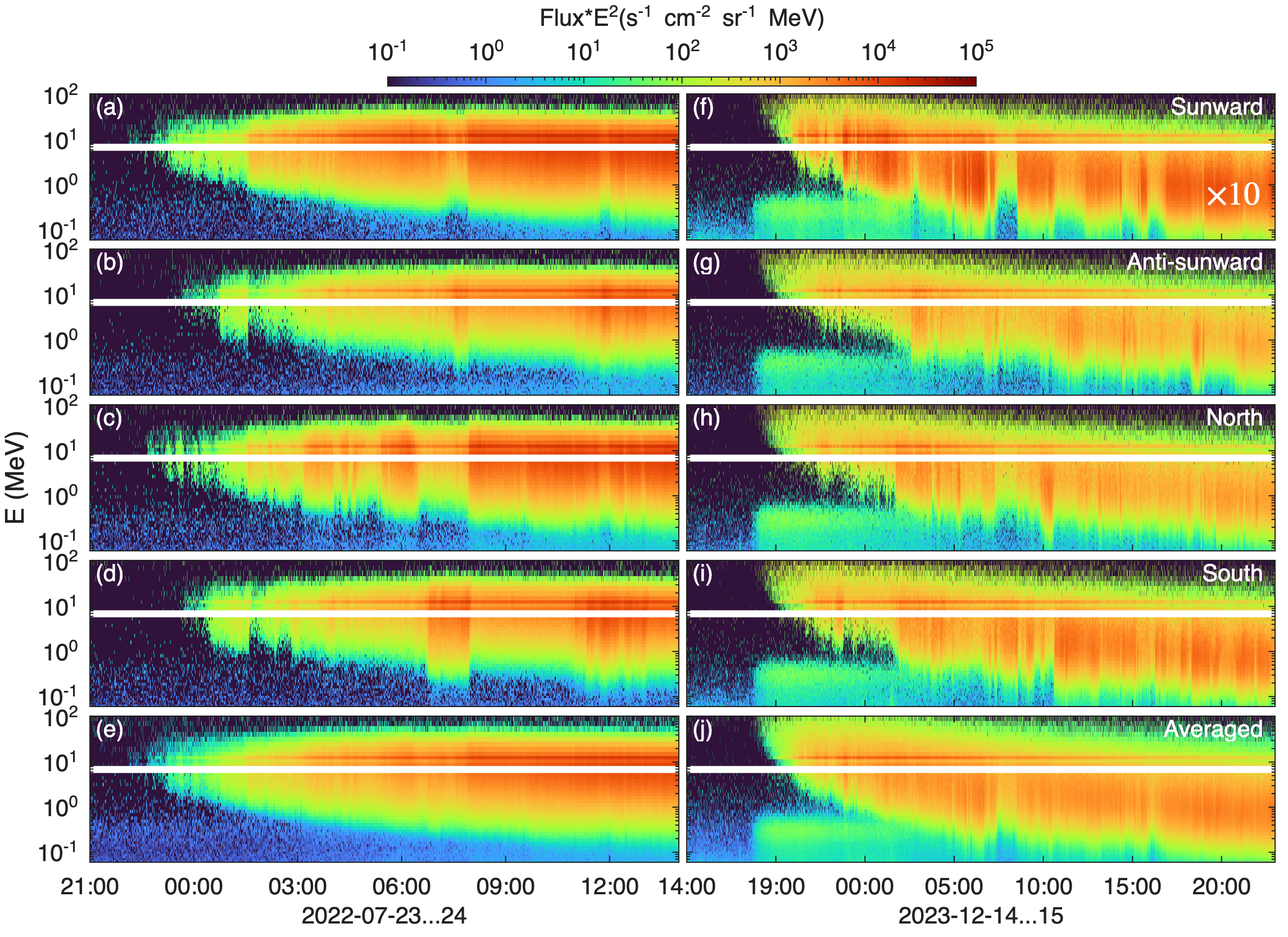}
\caption{Energy spectrograms observed by SolO/EPD for two intervals: (a–e) 2022 July 23 21:00 UT to July 24 14:00 UT, and (f-j) from 2023 December 14 16:00 UT to December 15 23:00 UT. In both cases, the top four panels show observations from the HET and EPT in four looking directions (sunward, anti-sunward, north, and south), and the lowest panel shows the averaged spectrogram. The flux intensities in (f-j) are scaled by a factor of 10 to enhance the visibility.} 
\label{Fig:overview}
\end{figure}

Figure \ref{Fig:overview} presents examples of SFDs detected in two SEP events. The left panel shows the event from 2022 July 23, 21:00 UT to July 24, 14:00 UT, and the right panel shows the event from 2023 December 14, 16:00 UT to December 15, 23:00 UT. During these intervals, SolO was located at heliocentric distances of 0.99 au and 0.91 au, respectively. From top to bottom, Figure \ref{Fig:overview} displays combined EPT and HET energy spectrograms in four looking directions (sunward, anti-sunward, north and south), followed by the flux averaged across all sensors (bottom panel). To improve visibility, the intensities in the second event have been scaled by a factor of 10. A series of SFDs are observed, characterized by significant variations in the time-intensity profiles. However, unlike dropout events where the flux decreases uniformly in all directions, the averaged intensities during SFDs (e.g., Figure \ref{Fig:overview} (e) and (j)) show insignificant net changes comparing to individual looking directions. Instead, SEP fluxes exhibit strong anisotropies: a drop in one or more directions (e.g., Figure \ref{Fig:overview} (a) and (c) between 6:00–9:00 UT) is typically accompanied by a simultaneous increase in others (e.g., Figure \ref{Fig:overview} (b) and (d) during the same period). Therefore, these variations are primarily due to a sudden change in the SEP streaming direction rather than a true depletion in overall flux intensity.

\begin{figure}[ht]
\centering
\includegraphics[width=0.8\textwidth]{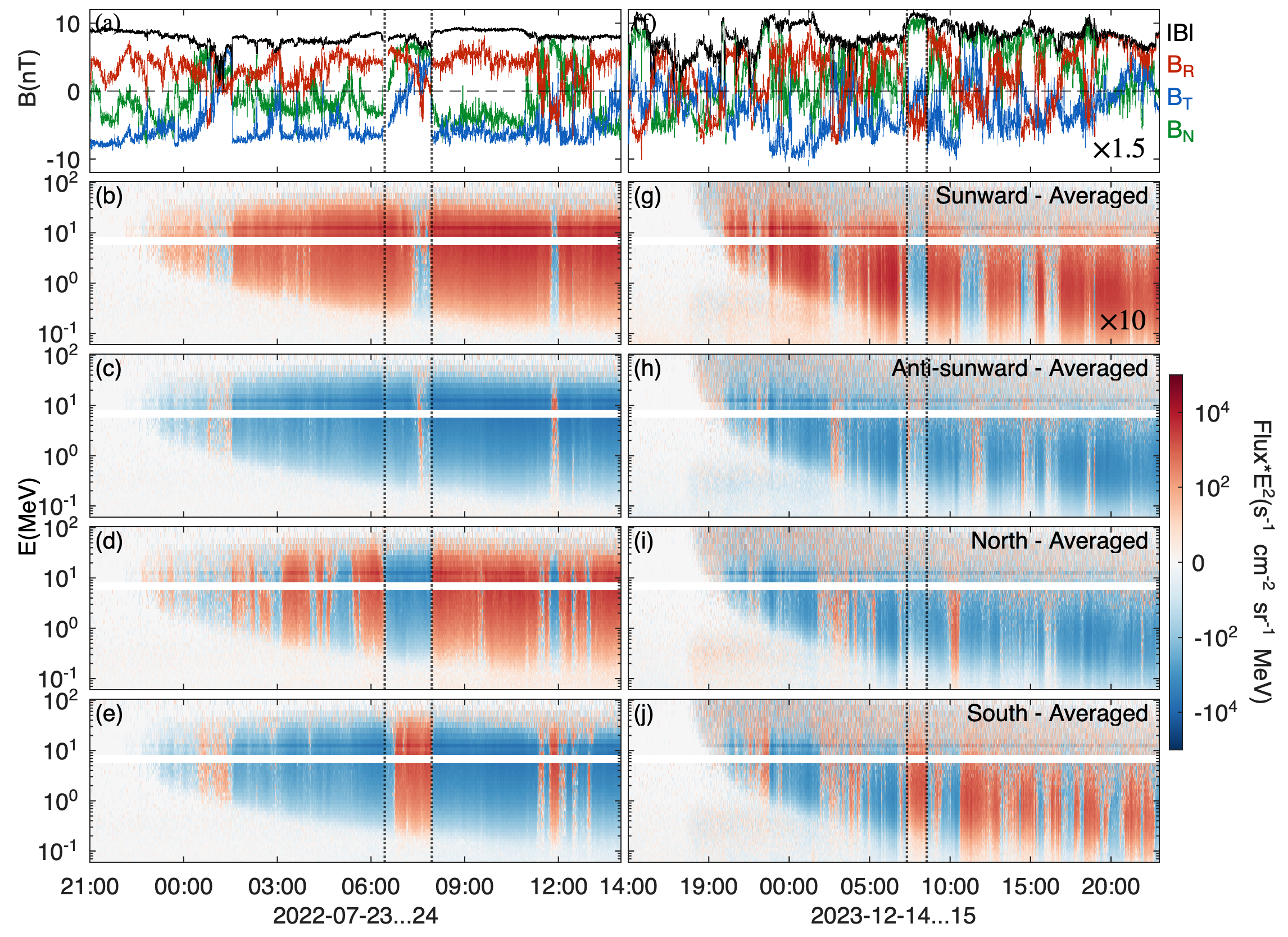}
\caption{Magnetic field components in RTN coordinates and energy spectrograms in four looking directions (with the averaged value subtracted) of two SEP events in Figure (\ref{Fig:overview}). The vertical dotted lines outlined the selected SFDs intervals in each event. The magnetic field and flux intensities in the right panels are scaled by factors of 1.5 and 10, respectively.} 
\label{fig:subtracted-flux}
\end{figure}

To better visualize these structures, we plot energy spectrograms from the four viewing directions after subtracting the averaged value. Figure \ref{fig:subtracted-flux} presents the magnetic field magnitude and vector components in the spacecraft-centered RTN coordinate system, along with the subtracted spectrograms for both events. The magnetic field strength and spectrogram intensity in the second event (right panels) have been scaled by factors of 1.5 and 10, respectively. A bi-symmetric logarithmic transformation \citep{Webber2013} is applied to avoid singularities that arise from zero values when displaying the spectrograms on a logarithmic color scale. Vertical dotted lines mark the selected intervals corresponding to SFD structures. Sharp and dispersionless (vertical) boundaries are observed at the edges of SFDs in the spectrograms, and a magnetic field rotation is typically associated with these structures (Figure \ref{fig:subtracted-flux} (a) and (f)). These sharp boundaries indicate that SEPs across all energies are well confined within these magnetic structures. As the structures convect with the solar wind past the observer, abrupt transitions in the SEP streaming direction are observed. 

\begin{figure}[ht]
\centering
\includegraphics[width=0.8\textwidth]{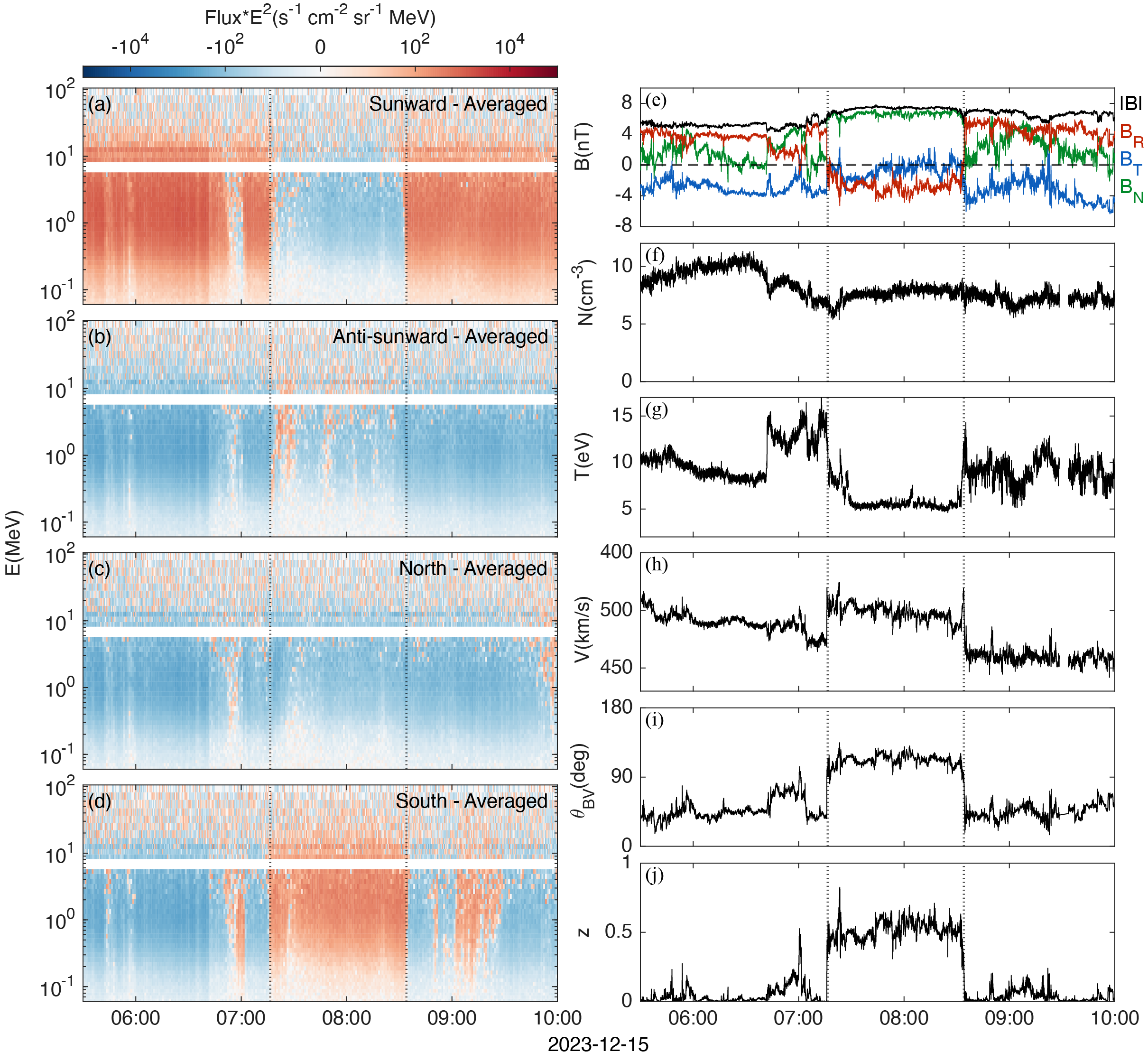}
\caption{SEP and solar wind observations from 05:30 to 10:00 UT on 15 December 2023. The SFD interval is marked by vertical dotted lines. (a–d) Background-subtracted energy spectrograms in four looking directions. (e) Magnetic field components in RTN coordinates. (f) Proton number density. (g) Proton temperature. (h) Bulk solar wind speed. (i)The angle between the magnetic field and the ambient solar wind flow. (j) Normalized magnetic field deflection.} 
\label{fig:zoom-in}
\end{figure}

Figure \ref{fig:zoom-in} provides more detailed observations of the selected SFD (marked by vertical dotted lines) in the second SEP event. The left panels show the background-subtracted SEP fluxes from 05:30 to 10:00 UT on 15 December 2023. Before the SFD, the SEP flux flows primarily along the IMF outward from the Sun; inside the structure, it deflects northward (being observed mostly in the south aperture). The right panels of Figure \ref{fig:zoom-in} show the corresponding solar wind properties over the same interval. The solar wind velocity remains nearly radial throughout this interval, so we plot only the bulk speed for clarity. The magnetic field lies generally along the nominal Parker spiral in the R–T plane before the SFD crossing, but rotates northward inside the SFD with little change in magnitude (see Figure \ref{fig:zoom-in} (e)). Proton density gradually decreases prior to the SFD but remains relatively unchanged afterward. In Figures \ref{fig:zoom-in} (g) and (h), proton temperature shows a slight decrease, while the bulk solar wind speed exhibits a modest increase, suggesting that the observer may have entered a distinct solar wind stream inside the SFD. Figure \ref{fig:zoom-in} (i) displays the angle between the magnetic field and the solar wind velocity, revealing a deflection of over 90 degrees.  We note that the radial component of the magnetic field reverses from positive to negative—a feature similar to the magnetic switchbacks recently observed near the Sun \citep[see review by][ and references therein]{Wyper2026}. To quantify the deflection, Figure \ref{fig:zoom-in} (j) presents the normalized magnetic field deflection $z = (1-\cos \alpha)/2$, where $\alpha$ is the deflection angle relative to the background field \citep{DudokdeWit2020}.

\begin{figure}[ht]
\centering
\includegraphics[width=0.8\textwidth]{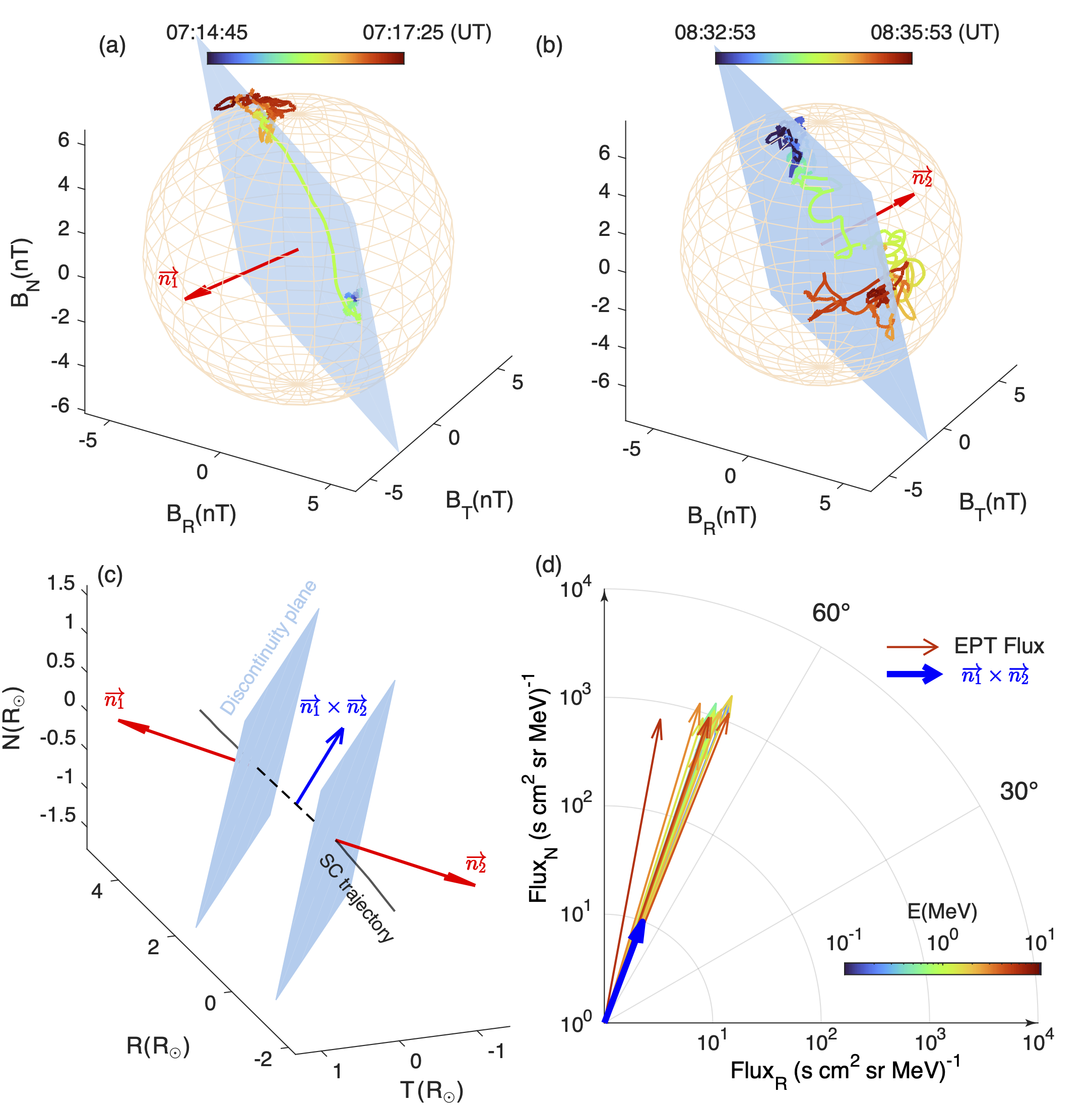}
\caption{Discontinuity analysis across the SFD boundaries on 15 December 2023. (a-b) Magnetic hodograms across the leading (07:16:05 UT) and trailing (08:34:23 UT) boundaries, where the color bar denotes the progression of time. The blue plane and red arrow indicate the discontinuity plane and its normal vector obtained from the SVD method. The orange sphere (radius = $\mathrm{|B|}$) represents the averaged field magnitude during the transition. (c) Orientation of the discontinuity planes relative to the SolO trajectory (black lines) during the SFD crossing. The blue arrow indicate the direction of the magnetic flux tube axis. (d) Directions and magnitudes of SEP fluxes from 100 keV to 10 MeV relative to the flux tube axis in the R-N plane. } 
\label{fig:boundary}
\end{figure}

We next analyze the magnetic fields at the leading (07:16:05 UT) and trailing (08:34:23 UT) boundaries of the SFD in the second event of 2023, as shown in Figures \ref{fig:boundary} (a) and (b). The temporal evolution of the magnetic field is represented by colored curves in the RTN coordinate system, with the color bar denoting the progression of time. The orange sphere represents the average magnetic field magnitude over this interval (radius = $|B|$). To determine the plane of the discontinuity, we apply the singular value decomposition (SVD) method \citep{Golub2013}. The plane that best captures the maximum rotation of the magnetic field vectors is shown in blue, from which we derive the normal directions of the discontinuity planes (red arrows): $\mathbf{n_1} = -[0.60\ 0.76\ 0.23]$ and $\mathbf{n_2} = [0.51\ 0.82\ 0.27]$. The signs of the normal vectors are chosen to point outward from the SFD. Following \citet{Bizien2023}, we determine the type of discontinuity in this super-Alfv\'enic solar wind by computing the ratio between the jump in the tangential component of the flow velocity and the jump in the tangential component of the Alfvén velocity: $\mathrm{R_{VB} = \Delta V_t / \Delta V_{At} = \sqrt{4 \pi \rho} \Delta V_t / \Delta B_t}$. The calculated $\mathrm{R_{VB}}$ values for the discontinuity planes in Figures \ref{fig:boundary} (a) and (b) are 0.76 and 0.65, respectively, indicating that the planes are more likely to be tangential discontinuities rather than rotational discontinuities ($\mathrm{R_{VB}} \sim 1$). Given the strong confinement of SEPs within this structure, this suggests that the magnetic structure could be a magnetic flux tube. Figure \ref{fig:boundary} (c) shows the orientation of the discontinuity planes (blue planes) relative to the SolO trajectory (black lines) during the SFD crossing, accounting for the solar wind velocity. Assuming the flux tube is cylindrical, its axis can be obtained as $\mathbf{n_1} \times \mathbf{n_2}$ (blue arrow). We then project the SEP fluxes from the sunward, anti-sunward, south and north directions to the RTN coordinate system. Figure \ref{fig:boundary} (d) presents the orientation of the flux tube axis $\mathbf{n_1} \times \mathbf{n_2}$ (thick blue arrow) and SEP fluxes from 100 keV to 10 MeV in the R-N plane. Colors indicate energy, and vector lengths indicate flux magnitude. The SEP fluxes align well with the flux tube direction, suggesting that within the structure, SEPs stream along the flux tube. (Note that the EPT-HET unit does not measure SEP flux from the azimuthal direction; consequently, the flux in the T direction is not compared here.) These observations indicate that the deflection of SEP fluxes is primarily due to the observer encountering a magnetic flux tube oriented differently from the background magnetic field.

%% file: Tex5_Conclusions.tex
\section{Discussion and Conclusions} \label{sec:Conclusions}

This study presents observational evidence of a distinct class of SEP variations—termed SFD—using energetic particle data from SolO. The small net change in the averaged SEP fluxes (Figure \ref{Fig:overview}(e)) suggests that the associated flux tubes are connected to SEP sources near the solar surface and that SEPs inside and outside the structures travel along similar path lengths to the observer. These results further indicate that some coherent structures observed in the solar wind—such as flux tubes and current sheets—may originate from the solar surface rather than being generated locally. However, the mechanism that generates these flux tubes is still under debate. They could originate either from interchange reconnection near the solar surface \citep{Drake2006,Daughton2011} or from the boundaries of granules and supergranules that connect from the photosphere into the solar wind—i.e., the spaghetti model \citep{Bruno2001,Borovsky2008}. A statistical study will be performed in a follow-up investigation to determine their origin.

\begin{figure}[ht]
\centering
\includegraphics[width=0.5\textwidth]{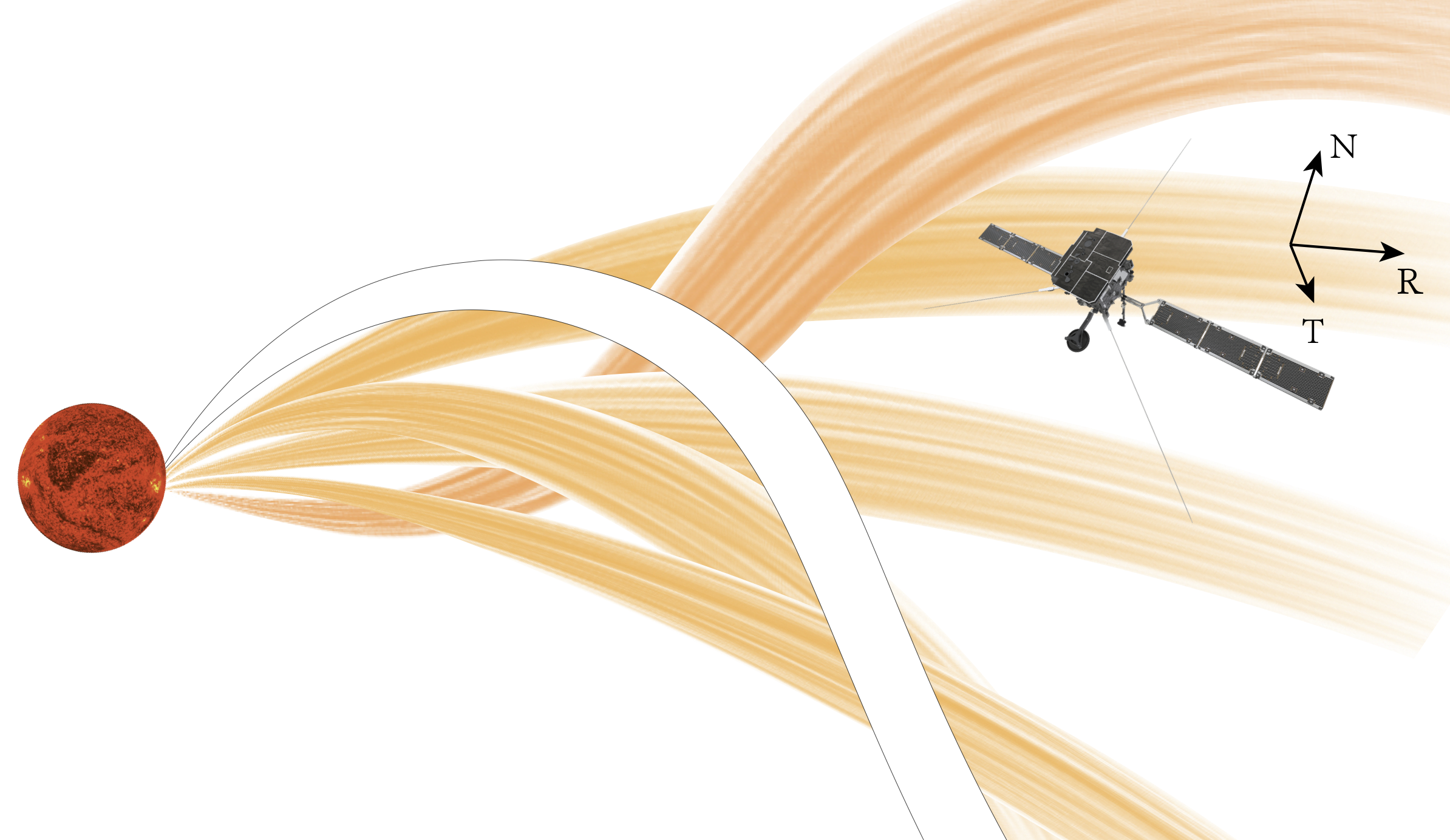}
\caption{Schematic illustrating SolO traversing meandering magnetic flux tubes (not to scale). Tubes filled with SEPs are shown in yellow, while the empty tube is left uncolored.} 
\label{fig:cartoon}
\end{figure}

These structures are carried outward into the heliosphere by the solar wind and may tangle up in interplanetary space (Figure \ref{fig:cartoon}). SEPs typically stream along the magnetic flux tube from the eruption site out to 1 au and beyond. Thus, SEP fluxes exhibit a deflection when an observer encounters a flux tube whose orientation differs from that of the background magnetic field. If the observer enters a flux tube devoid of SEPs—i.e., one not connected to the source region—a regular dropout will be detected. These flux tube structures appear to be steady and prevalent features of the solar wind, as they can be identified in both impulsive and gradual SEP events \citep[e.g.,][]{Ho2022,Wimmer-Schweingruber2023,Wimmer-Schweingruber2026}. SEPs associated with SFDs exhibit strong anisotropy, suggesting that particles undergo less scattering within these structures. The sharp and dispersionless boundaries of SFDs imply that little particle transport occurs across the boundary over a broad range of energies. Since the EPD does not cover $4\pi$ steradians, variations in the SEP flux across its four viewing directions reduce the observed flux magnitude, causing SFDs to resemble dropouts (see Figure \ref{Fig:overview}(j)). Consequently, EPD can only provide a lower limit on the number of SFDs. Moreover, given the limited number of SEP viewing directions (typically one or two) on previous spacecraft missions, some fraction of what were previously identified as dropouts may in fact be SFDs. These results reveal magnetic flux tubes as intrinsic structures of the solar wind, establishing SEP observations as a powerful tool for probing large-scale heliospheric architecture.

%% file: Tex6_Acknowledgments.tex
\begin{acknowledgments} \label{sec:Tex7_Acknowledgments}

 Solar Orbiter is a mission of international cooperation between ESA and NASA, operated by ESA. We acknowledge the energetic particle, solar wind plasma and magnetic field data from Solar Orbiter, generated and maintained by the EPD, SWA and MAG teams on the Solar Orbiter Archive (SOAR). Solar Orbiter post-launch work at JHU/APL is supported by NASA contract NNN06AA01C, at the Southwest Research Institute by NASA 80GFSC25CA035. XC, GL and SF acknowledge support at MUST through Science and Technology Development Fund (FDCT) of Macau grants 002/2024/SKL, 0008/2024/AKP, 0002/2025/AKP, 0006/2025/RIA1, 0137/2025/AFJ, 0022/2025/ASJ and Guangdong Basic and Applied Basic Research Foundation grant 2024A1515011994. XK acknowledges support by the National Natural Science Foundation of China (NNSFC) grants 42561160095 and 42574218. ZD and RFWS thanks the German Space Agency (DLR) for their unwavering support trough grant \#50OT2002.

\end{acknowledgments}